\documentclass[12pt,preprint]{aastex}

\begin{document}

\title{KELT: The Kilodegree Extremely Little Telescope}

\author{Joshua Pepper, Andrew Gould and  D.\ L.\ DePoy}

\affil{Department of Astronomy, Ohio State University, Columbus, Ohio 43210}
\email{pepper@astronomy.ohio-state.edu, gould@astronomy.ohio-state.edu, 
depoy@astronomy.ohio-state.edu}

\begin{abstract}
Transits of bright stars offer a unique opportunity to study detailed
properties of extrasolar planets that cannot be determined through
radial-velocity observations.  We propose a technique to find such systems
using all-sky small-aperture transit surveys.  The optimal telescope
design for finding transits of bright stars is a 5 cm ``telescope'' with a $4k
\times 4k$ camera.  We are currently building such a  system and expect to
detect $\sim 10$ bright star transits after one year of operation.
\end{abstract}

\maketitle

\hspace{1.5in}

\section{Introduction}

Transit surveys are widely believed to provide the best means to discover large number of extrasolar 
planets.  At the moment, all ongoing transit surveys are carried out in relatively narrow pencil beams.  
They make up for their small angular area with relatively deep exposures.  These surveys are potentially 
capable of establishing the frequency of planets in various 
environments, but they are unlikely to find the kinds of transits of bright stars that would be most 
useful for intensive follow-up analysis. Although some of the surveys of field stars are 
considered "wide field", their total survey areas are small compared to $4\pi$ sr.

\section{Optimal Telescope Design}

It can be demonstrated that the best way to find HD209458b - type transits is with an all-sky photometric 
survey.  According to Pepper, Gould \& DePoy \cite{PGD} (hereafter PGD), 
\begin{equation} \label{equ:dNt}
\frac{dN_t}{dM_V} = 5 \frac{\Omega}{4 \pi} F(M_V) \frac{f(M_V, r, a)}{0.75\%} 
\left(\frac{a}{a_0}\right)^{-5/2} \left(\frac{r}{r_0}\right)^6 \left(\frac{\gamma}{\gamma_0}\right)^{3/2} 
\left(\frac{\Delta \chi^{2}_{\rm min}}{36} \right)^{-3/2} 
\end{equation}
where $\gamma$ is the number of photons collected from a fiducial $V = 10$ mag star during the entire 
experiment, $\Delta \chi^{2}_{\rm min}$ is the minimum difference in $\chi^{2}$ between a transit 
lightcurve and a flat lightcurve required for candidate detection, and $F(M_V)$ is the function: 
\begin{equation} \label{equ:FMV}
F(M_{V}) = \left[ \frac{n(M_{V})}{n_{0}} \right] \left[ 
\frac{L(M_{V})}{L_{0}} \right]^{3/2} \left[ 
\frac{R(M_{V})}{R_{0}} \right]^{-7/2} 
\end{equation}
In equation (\ref{equ:dNt}) we have adopted  $\gamma_0 = 1.0 \times 10^7$, 
$a_0 = 10\,R_{\odot}$, $r_0 = 0.10\,R_{\odot}$, and in equation (\ref{equ:FMV}) we have made our evaluation at 
$M_V = 5$ mag (i.e. $R_0 = 0.97\,R_{\odot}$, $L_0 = 0.86\,L_\odot$, and $n_0 = 0.0025 \,{\rm pc}^{-3}$).  Note 
that $\gamma_0 = 1.0 \times 10^7$ 
corresponds to approximately 500 20-second exposures with a 5 cm telescope and a broadened (V+R) type filter 
for one $V = 10$ mag fiducial star. Since $N_t$ depends on the characteristics of the survey primarily 
through the total photon counts, $\gamma$, and only logarithmically (through $\Delta \chi^{2}_{\rm min}$) 
on the sampling strategy and the size of the explored parameter space (see PGD), telescope design must focus 
on maximizing $\gamma$, which is given by 
\begin{equation} \label{equ:KELT}
\gamma = \frac{K E L^{2} T}{\Omega F^2}
\end{equation} 
where $E$ is the fraction of the time actually spent exposing, $L$ is the linear size of the 
detector, $T$ is the duration of the experiment, $F$ is the focal ratio of the optics, and $K$ is a 
constant that depends on the telescope, filter, and detector throughput.  (For these calculations, we 
will assume $K = 40 \,{\rm e^- cm^{-2} s^{-1}}$, which is appropriate for a broad V+R filter and the fiducial 
$V = 10$ mag star.)  Interestingly, almost regardless of other characteristics of the system, the camera 
should be made as fast as possible.  We will adopt $F = 1.8$, beyond which it becomes substantially 
more difficult to design the optics. A more remarkable feature of equation (\ref{equ:KELT}) is that all 
explicit dependence on the size of the primary optic has vanished: a 1 cm telescope and an 8 m telescope 
would appear equally good.

PGD shows that two additional considerations (read-out time, scintillation noise) drive one 
toward smaller apertures, while two others (sky noise and focal-plane distortion) prohibit going 
beyond a certain minimal size. Combined, these four effects imply a sweet spot, 
\begin{equation} \label{equ:D}
D \approx (5 \rm cm) 10^{0.2(V_{\rm max} - 10)}
\end{equation}
where $V_{\rm max}$ is the targeted magnitude limit of the survey.

\section{The Kilo-square-degree Extremely Little Telescope (KELT)}

\begin{figure} \label{fig:curves}
  \includegraphics[height=.8\textheight]{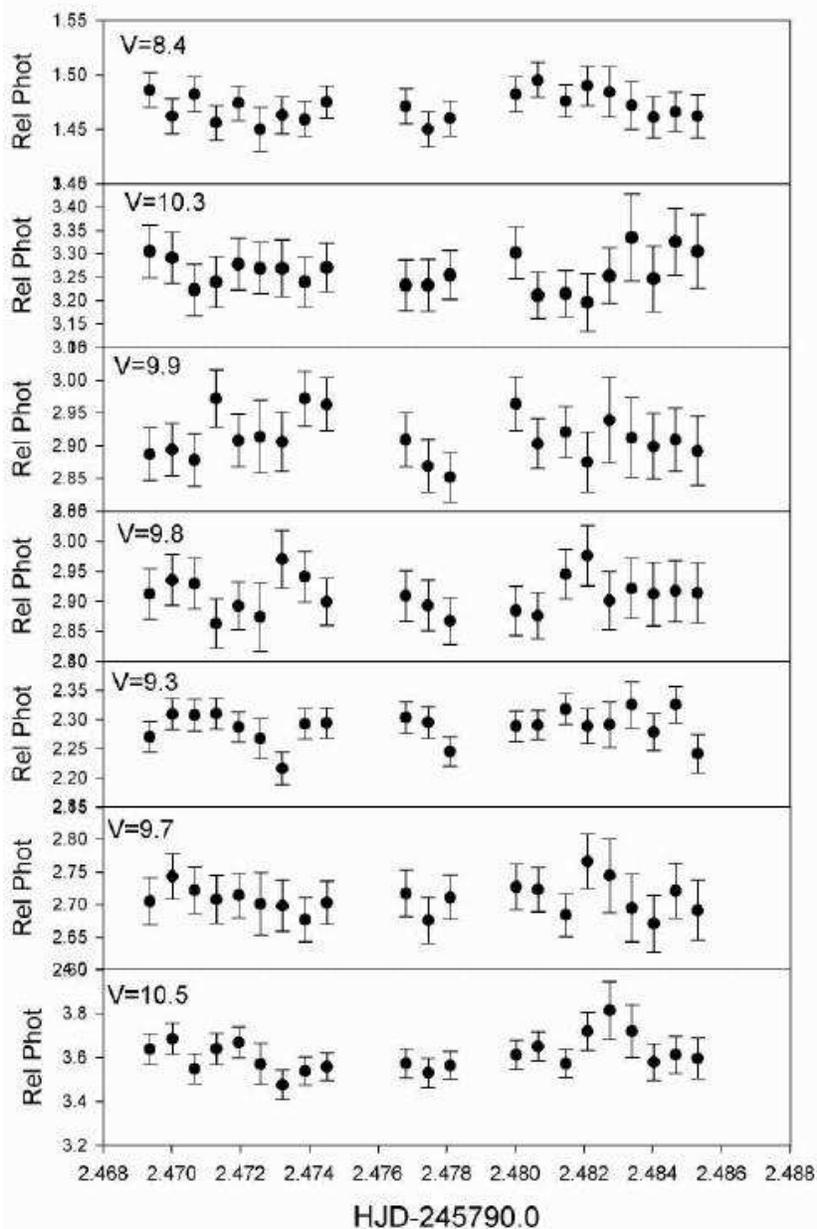}
  \caption{Relative light curves of 7 stars located near the center of the field.  This data was taken at a 
site in Ohio in June 2003.  The approximate visual magnitudes are given in each panel.  In 
general, we achieve 1-4\% photomet
ric accuracy for stars with $V=6-10$ mag.  These estimates are based on simple aperture photometry
 against a $V=19$ mag arcsec$^{-2}$ sky.   We expect a factor of at least 2-4 improvement using DoPhot
data reductions of photometry at Kitt Peak.}
\end{figure}

Motivated by the logic of the ``KELT equation'' (\ref{equ:KELT}), we have begun building the Kilodegree 
Extremely Little Telescope (KELT). As suggested by equation (\ref{equ:D}), KELT has a D = 4.2 cm, f/1.9 
lens (a Mamiya medium-format photographic lens) mounted on a $4096 \times 4096$ CCD with $9 \mu \rm m$ pixels 
(an Apogee Instruments AP16e camera using a Kodak KAF-16801E detector).  The images are Nyquist sampled 
over the entire field, as required to simultaneously minimize the problems of sky noise and intra-pixel 
variations.  Field tests show that the focal-plane distortions are manageable even in the corners of 
the $26^{\circ} \times 26^{\circ}$ field.  Test images were taken with this initial KELT system in 
June 2003 in Ohio. The images are roughly 
as expected for the optical performance of the Mamiya medium-format camera lens used. Figure \ref{fig:curves} 
shows representative light curves for several of the stars. 
In general, we achieve 1-4\% 
photometric accuracy for stars with $V = 6-10$ mag. These estimates are based on simple aperture 
photometry against a $V = 19$ mag arcsec$^{-2}$ sky, and we expect a factor of 2-4 improvement using 
DoPhot data reductions of photometry at Kitt Peak.

\subsection{Data Acquisition}

We plan to install KELT at a host site in New Mexico, where the telescope will be in an enclosure that will be 
opened every night. The telescope will execute a standard cycle of observations in fixed terrestrial (alt-az) 
coordinates, covering the areas of the sky within about $45^{\circ}$ of the zenith.  Each 30 s exposure will 
be tracked, and pointing to the next field will be executed during the 30 s read-out time.  The region 
within $45^{\circ}$ of zenith could be covered in 10 separate pointings, but the two most southerly ones will 
be duplicated and one of the two most northerly deleted in order to equalize sky coverage over the long 
term.  That is, each cycle will require 11 minutes, after which it will be repeated on the same section of 
the terrestrial sky (which has now moved $3.75^{\circ}$ to the east in celestial coordinates). In this way, the 
observations will (over the course of entire year) obtain roughly uniform coverage of about $2 \pi$ of 
the sky, roughly the northern hemisphere.

We anticipate $E = 50\% \times 20\% = 10\%$, where the first factor accounts for read-out time and the 
second for time lost to daylight, weather, and instrument problems.  We will employ a broad 
(V + R) filter for which we expect $K = K_0 =40 \,{\rm e^- cm^{-2} s^{-1}}$.  Hence, according to 
equation (\ref{equ:KELT}), 
during one year of observations, $\gamma = 7 \times 10^7$ photons will be collected from each $V = 10$ mag 
star.  That is $\gamma = 7 \times \gamma_0$.  Recall from equation (\ref{equ:dNt}) that $\gamma_0$  photons 
were required to probe $V_{\rm max} = 10$ mag stars for the transit of Jupiter-sized planets.

However, equation (\ref{equ:dNt}) is based on source-photon statistics alone. According to PGD, the photon 
requirements should be increased by a factor $1^2 + (4.2/5)^{2/3} + 0.75(4.2/5)^{-2}(1.9/1.8)^{-2} = 2.8$.  Hence, 
in 1 year, there is a "margin of safety" (to allow for unanticipated and/or unmodeled problems) of a factor 2.5 
in photon counts, corresponding to a factor 1.6 in S/N.  For three years of operation, the margin of safety is 
about a factor 2.7 in S/N.  We believe that this is adequate to ensure success.

\subsection{Data Analysis}

Photometry of each image will be carried out using a modified DoPhot package that is already 
well tested on microlensing data.  The positions and magnitudes of all target stars brighter than 
10 mag (and indeed several mag fainter) are already known from Tycho-2. Where necessary, 
this can be supplemented with USNO-A data to fainter mags. Thus, DoPhot can operate in its 
more efficient fixed-position mode.  The main modification required will be to take account of 
clouds.  In traditional "small-field" ($< 30$ arcmin) monitoring, accurate relative photometry is 
not seriously affected by clouds because the clouds dim all stars by approximately the same 
amount.  This will certainly not be the case for our $\sim 25^{\circ}$ field.  We will attempt to mitigate this 
problem by measuring star brightnesses only relative to nearby stars.  In the final analysis, 
however, we expect to lose more time to clouds than is lost in smaller-field monitoring.

Identification of transit candidates should be fairly straightforward. At $V = 10$ mag, there will be 
1\% errors (allowing for both scintillation and sky noise).  Of course, this is in itself not good 
enough to plausibly identify the 1\% transits due to Jupiter-sized planets.  However, a 2-hr transit 
should yield 11 such measurements and hence a $3\sigma$ signal.  Hence, it is not necessary to test all 
possible folds of the lightcurve to search the very large (period/phase) parameter space: one can 
focus first on the restricted space consistent with all subsets of the $2\sigma$ individual transit 
detections.

In the Northern sky, there are approximately 153,000 stars with $V \leq 10$ mag. Gould \& Morgan \cite{GM} showed that 
about 104,000 of these can be identified as being significantly evolved or 
early type (and so too large to be useful for planetary transits) using a Tycho-2 reduced proper 
motion (RPM) diagram.  Rejection of evolved stars will not only speed up the data processing, it 
will also remove one of the major sources of false candidates, namely K giants blended with 
eclipsing binaries (whether forming an associated triple or not). Vetting of the remaining 
candidates by follow up photometry during predicted transits using larger telescopes and by RV 
measurements will be much easier than for transit candidates from other surveys, simply because 
the KELT candidates will be extremely bright.

\hspace{1.5in}

\end{document}